\documentclass[a4paper,12pt]{article}

\usepackage{amsmath,amssymb,epsfig}

\def\appendix{{\newpage\section*{Appendix}}\let\appendix\section%
        {\setcounter{section}{0}
        \gdef\thesection{\Alph{section}}}\section}

\renewcommand{\a}{\alpha}
\renewcommand{\b}{\beta}

\renewcommand{\c}{\chi}
\renewcommand{\d}{\delta}

\renewcommand{\l}{\lambda}
\newcommand{\m}{\mu}

\newcommand{\p}{\pi}
\newcommand{\q}{\theta}

\newcommand{\s}{\sigma}
\renewcommand{\t}{\tau}

\newcommand{\lra}{\longrightarrow}

\newcommand{\cS}{\mathcal{S}}
\newcommand{\cO}{\mathcal{O}}

\newcommand{\ZZ}{\mathbb{Z}}

\newcommand{\CC}{\mathbb{C}}
\newcommand{\PP}{\mathbb{P}}

\newcommand{\beq}{\begin{equation}}
\newcommand{\eeq}{\end{equation}}

\DeclareMathOperator{\ch}{ch}
\DeclareMathOperator{\Ch}{Ch}
\DeclareMathOperator{\tr}{Tr}
\DeclareMathOperator{\rk}{rank}
\DeclareMathOperator{\Ahat}{\widehat{A}}
\DeclareMathOperator{\Td}{Td}
\DeclareMathOperator{\supp}{supp}
\DeclareMathOperator{\coker}{coker}
\DeclareMathOperator{\Pic}{Pic}

\begin{document}

\begin{titlepage}

\begin{flushright}
CERN-TH/2000-174\\
hep-th/0009112
\end{flushright}
%\vspace{12 mm}
%\vfil\vfil
\vskip 3cm

\begin{center}
{\Large {\bf  Brane-Antibrane Systems on  Calabi-Yau Spaces}\\} 

\vspace*{15mm}
\vspace*{1mm}
Yaron Oz$^a$, Tony Pantev$^b$ and Daniel Waldram$^a$

\vspace*{1cm} 

\textit{$^a$ Theory Division, CERN \\ CH-1211, Geneva  23, Switzerland}\\

\vspace*{.5cm}

\textit{$^b$ Department of Mathematics, University of Pennsylvania,\\
  Philadelphia, PA 19104--6395 USA}\\ 

\vspace*{.5cm}
%\maketitle
\end{center}

\begin{abstract}
We propose a correspondence between brane-antibrane systems and
stable triples $(E_1,E_2,T)$, where $E_1,E_2$ are holomorphic vector
bundles and the tachyon $T$ is a map between them. We demonstrate that,
under the assumption of holomorphicity, the brane-antibrane field
equations reduce to a set of vortex equations, which are equivalent to
the mathematical notion of stability of the triple. We discuss some
examples and show that the theory of stable triples suggests a
new notion 
of BPS bound states  and stability, and curious
relations between brane-antibrane configurations and wrapped branes in
higher dimensions.

\phantom{\cite{1,Witten,horava,sentach,lerda,numerics,hori1,chi,ms,2,bmo}}

\phantom{\cite{d1,d2,d2,d4,d5,d6,d7,d8,d9,d10,d11,d12,d13,d14}}

\end{abstract}
\vskip 2.5cm

September  2000
\end{titlepage}

\newpage

%%%%%%%%%%%%%%%%%%%%%%%%%%%%%%%%%%%%%%%%%%%%%%%%%%%%%%%%%%%%%%%%%%%%%%%%%%

\section{Introduction}

Systems of non-BPS brane configurations have been extensively studied
recently (for a review see \cite{review}).
A basic non-BPS  system is the coincident brane-antibrane
configuration, which is not stable.
It has a tachyon on the world-volume of the branes that 
arises from the open string stretched between the branes and the
antibranes, and it is charged under the world-volume gauge groups.  The
decay of the system can be seen, from various viewpoints  \cite{1,bmo}, 
by the tachyon rolling down to the
minimum of its potential. Upon tachyon condensation one can end up
with lower dimensional BPS branes, if the original brane-antibrane
system contained the corresponding charges. 

In another line of research, much progress has been achieved
in the study of BPS branes at arbitrary points in the moduli space
of Type II string theory compactified on Calabi-Yau spaces \cite{d1,d14}.
In particular a new concept of stability has been introduced \cite{d10}.

This work will relate to these two lines of research. We will
propose a correspondence between brane-antibrane systems and stable
triples $(E_1,E_2,T)$, where $E_1,E_2$ are holomorphic vector bundles
and the tachyon, $T$, is a map between them. We 
demonstrate that, under the assumption of holomorphicity, the
brane-antibrane field equations reduce to a set of vortex
equations. The latter are equivalent to the topological notion of
stability of the triple $(E_1,E_2,T)$. 
This is quite analogous to the case of a single vector bundle where
solutions of Hermitian Yang-Mills equations correspond to stable
holomorphic bundles. We discuss some examples and
show that the theory of stable triples suggests a new notion of BPS
bound states upon tachyon condensation, and curious relations
between brane-antibrane configurations and wrapped branes in higher
dimensions. 

This paper is organized as follows. In section 2 we will propose the
correspondence, and show the equivalence between
the (holomorphic) field equations of the brane-antibrane system and 
the vortex equations. We will show that such configurations saturate a
Bogomol'nyi bound and demonstrate the relation by some examples. 
In section 3 we will relate our description of BPS D-branes as stable
triples to existing descriptions of BPS states. In particular, we will
show that the theory of stable triples suggests a new notion of BPS
bound states and stability. This may also allow new BPS configurations
that arise upon tachyon condensation. In section 4 we discuss some
generalizations, and will see that the theory of stable triples
suggests some curious relations between brane-antibrane configurations
and wrapped branes in higher dimensions.

\section{Tachyon condensation and stable triples}

In this section we will propose a correspondence between systems of  
coinciding branes and antibranes wrapping a manifold $X$
and stable triples $(E_1,E_2,T)$. $E_1$ and $E_2$ are holomorphic
vector bundles on $X$. Physically, they correspond to the branes and
antibranes respectively. $T$ is a homomorphism between the vector
bundles $T:E_2 \rightarrow E_1$. It is the tachyon field
that arises from the open string stretched between the branes and the
antibranes\footnote{To match the math literature we think of $T$ as a
map from antibranes to branes. The conjugate field $\bar{T}$ is the
map in the other direction.}. With a particular holomorphic ansatz, we
will recast the field equations of the brane-antibrane system as a set
of vortex equations. We argue that solutions of the vortex equations
represent BPS configurations. Such solutions correspond to a
mathematical construction of stable triples on $X$ \cite{Brad}. With
this correspondence the analysis of tachyon condensation leading to
BPS branes will be replaced by a stability analysis.

\subsection{Tachyon condensation and triples}

Let us now briefly review the construction of D-branes from
coincident brane-antibrane configurations \cite{Witten}. We will
consider the particular case of Type IIA string theory compactified on
Calabi--Yau manifold $X$ of complex dimension $d$. 

A configuration of $n$ branes wrapping the Calabi--Yau manifold is
described by a $U(n)$ vector bundle $E$ on $X$. This carries charges
for various RR fields which, as shown in \cite{hm,ghm}, takes the
form 
\begin{equation}
   Q = \ch(E)\sqrt{\Ahat(X)} = \ch(E)\sqrt{\Td(X)}\ ,
\label{ve}
\end{equation}
which is an element of the cohomology $H^*(X,\ZZ)$ known as the Mukai
vector. For the equality in (\ref{ve}) we used the fact that
$\Td(X)$ on a Calabi--Yau manifold is equal to the
A-roof genus $\Ahat(X)$. 
In this expression $\ch(E)$ is the Chern character of the
vector bundle $E$  
\begin{equation}
   \ch(E) = \tr \exp\left[\frac{F}{2\pi}\right] \ ,
\label{che}
\end{equation}
where $F$ is the field strength of the gauge field on the brane.
It has an expansion in terms of the
Chern classes  
\begin{equation}
   \ch(E) = n + c_1(E) + \frac{1}{2}(c_1^2(E) - 2c_2(E)) 
      + \dots \ . 
\label{ee}
\end{equation}
The 
A-roof genus $\Ahat(X)$ and has an expansion in terms of the
Pontrjagin classes 
\begin{equation}
   \Ahat(X) = 1 - \frac{p_1(X)}{24} + \dots \ .
\label{eA}
\end{equation}

Now suppose we have a configuration of $n_1$ branes wrapped on $X$,
together with $n_2$ antibranes. In general, the configuration 
is described by specifying a $U(n_1)$ vector bundle $E_1$ on $X$ for
the branes together with a second $U(n_2)$ bundle $E_2$ for the
antibranes. The net D-brane charge is then the difference of the Mukai
vectors for the two bundles
\begin{equation}
   Q = Q_1 - Q_2 =
   \left(\ch(E_1)-\ch(E_2)\right)\sqrt{\Ahat(X)} \ .
\label{eq:Q}
\end{equation}
In general, we would expect the antibranes to annihilate against the
branes. However, if the bundles are different then there is a net
D-brane charge so the branes cannot completely annihilate and still
conserve charge.  

Since identical bundles can annihilate, adding
the same bundle to $E_1$ and $E_2$ gives the same physical
configuration. That is we should identify $(E_1\oplus V,E_2\oplus V)$ with
$(E_1,E_2)$, which is the equivalence class identification made in
K-theory. In K-theory one identifies the Chern class of
the pair $(E_1,E_2)$ with the difference of Chern classes
$\ch(E_1)-\ch(E_2)$, thus providing the map from K-theory to the
Mukai charge~\eqref{eq:Q}. In fact, a D-brane charge is more accurately
measured by the K-theory class rather than the cohomological Mukai
charge. In particular, K-theory includes more information than the
Chern classes themselves. For instance, Chern classes miss torsion.

Physically, the annihilation happens because there is a tachyonic mode
$T$ in the open string connecting the branes and antibranes. The
tachyon potential has a minimum away from zero. However, if the
bundles $E_1$ and $E_2$ are different, there is a topological
obstruction to the tachyon being at the minimum everywhere on $X$. The
tachyon $T$ transforms in the fundamental representation of each
bundle $(\mathbf{n_2},\mathbf{\bar{n}_1})$, and it must respect the
twisting of each bundle. In general, even if $n_1=n_2$ if the bundles
are different, it cannot do so and remain everywhere at the minimum of
the tachyon potential. Instead it must be zero on some sub-manifold
$C$ of $X$. There is a vortex solution representing a
lower-dimensional brane localized on $C$. In particular, all the lower
dimensional branes can be built out of D9-branes in this way
\cite{Witten}.

To specify a general D-brane configuration we need to
specify the bundles $E_1$ and $E_2$ together with the condensed
tachyon $T$. Since $T$ is in the bi-fundamental representation, it
represents a map between the bundles. Thus the full information is the
triple $(E_1,E_2,T)$ giving the complex 
\begin{equation}
   E_2 \stackrel{T}{\longrightarrow} E_1 \ .
\label{eq:triple}
\end{equation}
As we have noted, the D-brane charges are characterized by the
K-theory class of the pair $(E_1,E_2)$. However for a given D-brane
charge there is generically a moduli space of different D-brane
states. It is natural to ask what characterizes these distinct
D-brane states. In general, this should be some equivalence
class of triples $(E_1,E_2,T)$, giving a finer classification than
simply the K-theory class. In this paper, we will consider BPS
configurations. We will see that this implies that the bundles and
maps are holomorphic. A possible equivalence class has then been
proposed in \cite{sharpe}. The suggestion is that for holomorphic
bundles $E_1$ and $E_2$, we should identify triples in the same
derived category \cite{GM,thomas}, which essentially means considering
complexes of bundles of the form~\eqref{eq:triple}, modulo exact
sequences.  Note, however, that such an identification is not suitable
for measuring stability, as the notion of stability is intrinsic only
for Abelian categories, while the derived category is just additive.

\subsection{The vortex equations}

Consider the low-energy effective action of the
world-volume theory of a  configuration of coinciding $n_1$
branes and $n_2$ antibranes wrapping a manifold $X$
\begin{equation}
   S = \int_X \left[ \frac{1}{4}\tr_1 F_1^2 + \frac{1}{4}\tr_2 F_2^2 
          + (DT)^a_{\ \bar{b}}(DT^*)^{\bar{b}}_{\ a}
          + \l \left(T^a_{\ \bar{b}}T^{*\bar{b}}_{\ \ b}
              -\a^2\d^{a}_{\ b}\right)^2 
          \right] \ .
\label{eq:action}
\end{equation}
Typically, we will take $n_1=n_2$, so only lower-dimensional branes
remain after condensation. There are higher order corrections to
(\ref{eq:action}), and in general one also expects, as for the
non-BPS branes~\cite{Berg}, the kinetic terms of the tachyon and the
gauge fields to depend on the tachyon background. Such corrections
modify the field equations and the precise description
of the tachyon rolling to the minimum of its potential \cite{wip}.
We expect, however, that it should not matter
for the topological construction of the lower-dimensional
branes upon the condensation of the tachyon. Here we think about the
lower-dimensional branes as the BPS branes or the
stable non-BPS ones (where the tachyonic mode is projected out).
Quantitative properties of the lower-dimensional
branes such as the size of vortex solutions will be modified, upon the
inclusion of the corrections.

We further assume the same gauge coupling for the two gauge groups
and rescaled the gauge and tachyon fields in  (\ref{eq:action}).
In~\eqref{eq:action}, $a$ is the index of the fundamental
representation of $E_1$ and $\bar{a}$ the anti-fundamental of $E_2$.
The parameter $\a^2$ in the tachyon potential is related to the value
of the tachyon field at the minimum of the potential
\begin{equation}
   \a^2 = \frac{1}{n_1}\tr(TT^*)|_{\text{minimum}} \ .
\label{a2}
\end{equation} 
Since $T$ has charge $\pm e$ under the gauge groups and its covariant
derivative is
\begin{equation}
   D_M T^a_{\ \bar{b}} = \partial_M T^a_{\ \bar{b}} 
        + ie(A^1_M)^a_{\ b} T^b_{\ \bar{b}}
        - ieT^a_{\ \bar{a}}(A^2_M)^{\bar{a}}_{\ \bar{b}} \ .
\label{eq:D}
\end{equation}
From now on we will suppress the indices and write, for instance
$DT=dT+ieA_1T-ieTA_2$. 

The equations of motion  read
\begin{equation}
\begin{aligned}
   D_1^M F^1_{MN} &= ie\left[ 
      T\left(D_NT^*\right) - \left(D_NT\right)T^* \right] \\
   D_2^M F^2_{MN} &= ie\left[ 
      T^*\left(D_NT\right) - \left(D_NT^*\right)T \right] \\
   D^2 T &= 2\l \left(TT^*T - \a^2T\right) \ , 
\end{aligned}
\label{eq:eom}
\end{equation}
where $D_1=d+ieA_1$ and $D_2=d+ieA_2$. 

We denote the K\"ahler metric on $X$ by $g_{m\bar{n}}$, where $m$ is a
holomorphic index and $\bar{n}$ an anti-holomorphic index. There is then
a set of equations which imply the equations of motion. They are,
first, that all the fields are holomorphic, namely
\begin{equation}
\begin{gathered}
   F^1_{mn} = F^1_{\bar{m}\bar{n}} = 0 \\
   F^2_{mn} = F^2_{\bar{m}\bar{n}} = 0 \\
   D_{\bar{m}}T = 0 \ .
\end{gathered}   
\label{eq:holo}
\end{equation}
Then in addition we have a Hermitian condition
\begin{equation}
\begin{aligned}
   ig^{m\bar{n}}F^1_{m\bar{n}} + eTT^* &= 2\p\t_1 I_1 \\
   ig^{m\bar{n}}F^2_{m\bar{n}} - eT^*T &= 2\p\t_2 I_2
\end{aligned}
\label{eq:vortex}
\end{equation}
where $I_{1},I_{2}$ are the identity matrices for the $E_1$ and $E_2$
bundles respectively. Together we shall call equations (\ref{eq:holo})
and (\ref{eq:vortex}) the {\it vortex
equations}. The important point, as we discuss in the next subsection,
is that solutions of the vortex equations (\ref{eq:holo}) and
(\ref{eq:vortex}) are in one-to-one correspondence with the
topological notion of stability of the triple $(E_1,E_2,T)$ \cite{Brad}.

To see that the vortex equations do indeed imply the equations of
motion (\ref{eq:eom}), consider first the consequences of the
holomorphic conditions~\eqref{eq:holo}. Recall that the Bianchi
identity on either $F^1$ or $F^2$ reads 
\begin{equation}
   D_m F_{n\bar{n}} + D_n F_{\bar{n}m} + D_{\bar{n}} F_{mn} = 0 \ .
\label{eq:BI}
\end{equation}
Using the holomorphicity condition and contracting with $g^{m\bar{n}}$
implies that
\begin{equation}
\begin{split}
   D^MF_{Mn} &= g^{m\bar{n}} D_m F_{\bar{n}n} \\
       &= - D_n \left( g^{m\bar{n}}F_{m\bar{n}} \right) \ .
\end{split}
\label{eq:DF}
\end{equation}
Since, in addition, holomorphicity implies $D_mT^*=0$, we have that
the $F^1$ and $F^2$ equations of motion read
\begin{equation}
\begin{aligned}
   D_n \left(ig^{m\bar{n}}F^1_{m\bar{n}}\right) 
       &= - e D_n(TT^*) \\
   D_n \left(ig^{m\bar{n}}F^2_{m\bar{n}}\right) 
       &= e D_n(T^*T) \ ,
\end{aligned}
\label{eq:holoF}
\end{equation}
which is just the derivative $D_n$ of the vortex
equations~\eqref{eq:vortex}. 

Now we turn to the tachyon equation of motion. We have, commuting the
$D$ derivatives:
\begin{equation}
\begin{split}
   D^2T &= g^{m\bar{n}}\left(D_mD_{\bar{n}}T + D_{\bar{n}}D_mT\right) \\
      &= g^{m\bar{n}}\left(
           2D_mD_{\bar{n}}T - ieF^1_{m\bar{n}}T + ieTF^2_{m\bar{n}}
           \right) \\
      &= -e\left(ig^{m\bar{n}}F^1_{m\bar{n}}\right)T 
           + eT\left(ig^{m\bar{n}}F^2_{m\bar{n}}\right) \ ,
\end{split}
\label{eq:D2T}
\end{equation}
where in the last line we use the holomorphic properties of $T$. Thus
the tachyon equation of motion now reads
\begin{equation}
   \left(ig^{m\bar{n}}F^1_{m\bar{n}}\right)T 
       - T\left(ig^{m\bar{n}}F^2_{m\bar{n}}\right)
       = -\frac{2\l}{e}\left( TT^*T - \a^2T \right) \ .
\label{eq:Teom}
\end{equation}
This corresponds to pre- and post-multiplying the vortex equations by
$T$ and taking the difference. However, this requires that the
parameters in the action are related so that
\begin{equation}
   \l = e^2 \ .
\label{eq:lm-e}
\end{equation}
We also get a relation between $\t_1$, $\t_2$ and $\a$
\begin{equation}
   e\a^2 = \p \left( \t_1 - \t_2 \right) \ .
\end{equation}
Note that the relation between $\l$ and $e$, together with the
assumption that the height of the tachyon potential is the tension of
the brane system ${\cal T}_p$, imply that the tachyon charge $e$ is related to
the value of the tachyon field at the minimum of its potential by
\begin{equation}
   e^2 \sim {\cal T}_p \left(
       \frac{1}{n_1}\left.\tr(TT^*)\right|_{\text{minimum}}
       \right)^{-2} \ .
\end{equation}

That such a relation holds in general can be established by using the
vortex equations to construct the known BPS brane solutions 
at the minimum of the tachyon potential, as we will do later. 
It also requires that the trace structure of the
potential has a particular form, in particular the quartic terms are
${\rm Tr} TT^*TT^*$ with no $({\rm Tr} TT^*)^2$
term~\cite{horava}. Note that requiring the action to be
formulated in terms of a superconnection following
Quillen~\cite{quillen,Witten,roep}, also gives $\l=e^2$ and the given
trace structure in the potential. In what follows we set $\l=e=1$ for
convenience. 

If we add that vortex equations, take a trace and
integrate over $X$, we find 
\begin{equation}
   \t_1 n_1 + \t_2 n_2 = \deg E_1 + \deg E_2 \ ,
\label{eq:tau2}
\end{equation}
where the degree of a vector bundle $\deg E$ is defined as
\begin{equation}
   \deg E = \frac{1}{V(d-1)!} \int c_1(E) \wedge J^{d-1} \ ,
\label{eq:degree}
\end{equation}
with $J$ the K\"ahler form on $X$ and $c_1(E)$ is the first Chern
class. Thus we see that $\t_1$ and $\t_2$ are completely determined by
the parameter $\a$ and the bundles $E_1$ and $E_2$. In particular
\begin{equation}
\begin{aligned}
   2\p\t_1 &= 2\pi\frac{\deg E_1 + \deg E_2}{n_1+n_2} 
      + \frac{2n_2}{n_1+n_2}\a^2 \\
   2\p\t_2 &= 2\pi\frac{\deg E_1 + \deg E_2}{n_1+n_2}
      - \frac{2n_1}{n_1+n_2}\a^2 \ .
\end{aligned}
\label{eq:t1t2}
\end{equation}

We expect that the solutions of the vortex equations are
supersymmetric BPS states. One way to establish this is
to analyze the supersymmetry directly. Another way, which we will
follow, is to show that these solution satisfy the Bogomol'nyi bound.
Separating into holomorphic and anti-holomorphic indices, integrating
by parts on the tachyon kinetic terms, commuting the derivatives, and
using the identity 
\begin{equation}
   \int_X \sqrt{g} \tr F_{m\bar{n}}F^{m\bar{n}} = 
        \int_X \sqrt{g} \tr \left( ig^{m\bar{n}}F_{m\bar{n}} \right)^2
        + \frac{1}{(d-2)!} \int_X \tr F \wedge F \wedge J^{d-2} \ ,
\end{equation}
one can rewrite the action as
\begin{equation}
\begin{split}
   S =& \int_X \left[ \frac{1}{2} \tr_1 F^1_{mn}F^{1\,mn}
         + \frac{1}{2} \tr_2 F^2_{mn}F^{2\,mn}
         + 2 g^{m\bar{n}} \tr_1 D_{\bar{n}}T D_mT^* 
         \right. \\ & \left.
         + \frac{1}{2}\tr_1\left(
              ig^{m\bar{n}}F^1_{m\bar{n}} + eTT^* - 2\p\t_1I_1
              \right)^2 
         + \frac{1}{2}\tr_2\left(
              ig^{m\bar{n}}F^2_{m\bar{n}} - eT^*T - 2\p\t_2I_2
              \right)^2 \right]
         \\ &
         + \textrm{topological terms} \ .
\end{split}
\label{eq:bog}
\end{equation}
The topological terms involve the degrees of $E_1$ and $E_2$ as well as
the second Chern characters
\begin{equation}
   \Ch_2(E_i,J) = 
      \frac{1}{V(d-2)!}\int_X \ch_2(E_i) \wedge J^{d-2} \ ,
\end{equation}
all of which are fixed once the $E_1$ and $E_2$ bundles are
chosen. Note that each non-topological term in~\eqref{eq:bog} is
positive or zero, so that the action is bounded from below by the value
of the topological terms. Comparing with equations~\eqref{eq:holo}
and~\eqref{eq:vortex}, we see that the bound is saturated if and only
if one satisfies the vortex equations. Since $X$ is Euclidean, the
action is the energy of the configuration. Thus solutions of the vortex
equations give the states of global minimum mass for a given set of
charges (fixed by the bundles $E_1 $ and $E_2$). Since in a
supersymmetric theory this bound is saturated only by supersymmetric
BPS states, this implies that solutions of the vortex equations are
indeed BPS states.  

We have shown that with the relations (\ref{eq:lm-e}) and
(\ref{eq:t1t2}) we have an interesting correspondence. Brane
configurations where all fields are holomorphic and that arise via the
process of tachyon condensation are described by solutions to the
vortex equations (\ref{eq:holo}) and (\ref{eq:vortex}). There is one
dimensionful parameter, $\a^2$, in the equations which is related to
the value of the tachyon at the minimum of the potential and so scales
as the string scale. We should emphasize that the holomorphicity conditions  
(\ref{eq:holo}) limit our discussion to tachyon condensation that
leads to BPS branes. In order to study stable non-BPS branes we would
have to relax these conditions.

\subsection{Stable triples}

A particularly useful property of the vortex equations is that
their solutions are in one-to-one correspondence with a topological notion
of stability of the triple $(E_1,E_2,T)$. Thus we can use stability to
analyze the existence of solutions on a general $X$, rather than
looking for solutions explicitly. 

The analogy here is to the Hermitian Yang--Mills equations (HYM)
describing the supersymmetric compactification of a single gauge
bundle $E$ on a Calabi--Yau manifold. 
They read 
\begin{equation}
   F_{mn} = F_{\bar{m}\bar{n}} = 0 \qquad 
   ig^{m\bar{n}}F_{m\bar{n}} = 2\p\t \ ,
\label{eq:HYM}
\end{equation}
which are a simple subset of the vortex equations. By
the Donaldson--Uhlenbeck--Yau theorem, solutions of the HYM
equations are in one-to-one correspondence with holomorphic vector
bundles of a particular type: those that are
poly-stable. This is defined as follows. Let the slope of a bundle be
given by
\begin{equation}
   \m(E) = \frac{\deg E}{\rk E} \ .
\label{eq:slope}
\end{equation}
A bundle $E$ is stable if for any non-trivial sub-bundle $E'\subset
E$, one has $\m(E')<\m(E)$. Poly-stability means that $E$ is the
direct sum of stable bundles each with the same slope. By this
theorem, an analytic problem --- solutions of the HYM equations --- is
a equivalent to a topological problem --- listing the stable bundles
--- and the latter problem is generally much easier to solve. We
should note that the stability problem is not quite topological: it
also depends on the choice of K\"ahler form $J$ via the definition of
$\deg E$. 

It turns out there that is an analogous notion of stability for a
triple $(E_1,E_2,T)$, such that there is a solution to the vortex
equations if and only if the triple is stable~\cite{Brad}. One first
needs to define what is meant by a sub-triple. We take the definition
that $(E'_1,E'_2,T')$ is a sub-triple if 
\begin{enumerate}
\item[(1)] $E_i'$ is a coherent sub-sheaf of $E_i$ with $i=1,2$,
\item[(2)] $T'$ is the restriction of $T$.
\end{enumerate}
Equivalently one requires that the diagram
\begin{equation}
   \begin{array}{ccc}
      E_2& \stackrel{T}{\lra} &  E_1 \\
      \uparrow&  &  \uparrow\\
      E_2'& \stackrel{T'}{\lra} &  E_1'
   \end{array}
\end{equation}
is commutative. Next one needs the analog of the $\mu$-slope $\m(E)$. With
$\sigma$ a real number, one defines the $\sigma$-slope of a triple
$(E_1,E_2,T)$ by
\begin{equation}
   \m_{\sigma}(T) = \frac{\deg(E_1\oplus E_2)+\sigma n_2}{n_1+n_2} \ .
\label{slope}
\end{equation}
A triple is then called $\sigma$-stable if for all nontrivial sub-triples 
$(E'_1,E'_2,T')$ we have  
\begin{equation}
   \m_{\sigma}(T') < \m_{\sigma}(T) \ .
\label{stable}
\end{equation}

The relation between solutions to the vortex equations (\ref{eq:holo})
and (\ref{eq:vortex}), and the $\sigma$-stability of the triple is for
$\sigma = \tau_1- \tau_2$. As seen, for instance, from (\ref{eq:t1t2}), 
for the brane-antibrane system it means
\begin{equation}
   \sigma \equiv \frac{\alpha^2}{\pi} \ .
\label{sigma}
\end{equation}
The BPS branes that arise upon the condensation of
the tachyon will be $\sigma$-stable with $\sigma$ given by the
relation (\ref{sigma}).
As will be very relevant later, since $\alpha$ goes like the string
scale $M_s$, the large volume limit corresponds to large $\sigma$.
This is the regime where we can trust the vortex equations to provide
an adequate description.
Note, in particular in the large $\sigma$ limit, the stability condition
(\ref{stable}) reads
\begin{equation}
   n_2 n_1' - n_2'n_1 > 0 \ ,
\label{theta}
\end{equation} 
where $n_i' = \rk E_i'$, for $i=1,2$, which is similar to a stability
condition on a quiver in the orbifold limit \cite{d2}\footnote{We
  thank M. Douglas for  valuable discussions on the large $\sigma$
  limit, and for pointing out the relation of (\ref{theta}) to the one
  that arises in the quiver analysis (see appendix A of (\cite{d3})).}.

\subsection{An example}

As a simple illustration of solutions to the
vortex equations and some of the discussion
to follow, let us consider how we can realize a D0-brane on
$\CC$ via the condensation of a D2-brane and an anti-D2-brane. 

The D2-branes will be realized as $U(1)$ bundles. For finite energy,
we require that the connection is pure gauge at infinity (thus we are
effectively considering bundles on $S^2=\PP^1$). To ensure that we
have a zero brane we need the difference of the bundle
charges~\eqref{eq:Q} to be one: 
\begin{equation}
   c_1(E_1) - c_1(E_2) = 1 \ .
\label{eq:DOcharge}
\end{equation}
For simplicity we can take $E_2$ to be trivial, while $E_1$
has $c_1(E_1)=1$\footnote{We expect the existence of some equivalence
relation between triples, such that different choices
of bundles satisfying (\ref{eq:DOcharge}) will lead to the
same BPS state.}. This means that $E_1$ has a non-trivial holonomy at
infinity. 

Now consider a solution of the vortex
equations~\eqref{eq:holo}. First $T$ must be holomorphic. This
implies that 
\begin{equation}
   \bar{\partial}T + iA^1_{\bar{z}}T - iA^2_{\bar{z}}T = 0 \ .
\label{eq:DT}
\end{equation}
Equation (\ref{eq:DT}) can be solved and gives 
\begin{equation}
\begin{split}
   A^1_z - A^2_z &= i \partial \ln T \\
      &= \partial\c + i\partial \ln f \ ,
\end{split}
\label{eq:Adiff}
\end{equation}
where we have written the tachyon as $T=fe^{i\c}$. Note that locally
gauge transformations can always remove $\c$. Since fields must
be pure gauge at infinity, we have $\ln f \to \text{const}$ at
infinity. The fact that $E_2$ is trivial and $E_1$ has $c_1(E_1)=1$
means that $A_2$ can be gauged to zero at infinity while $A_1$ can be
gauged to the form 
\begin{equation}
   A_1|_\infty = \partial \q \ ,
\label{eq:A1}
\end{equation}
where $z=re^{i\q}$, which is locally trivial but has global
holonomy. 

From this we see that we can gauge $T$ such that 
\begin{equation}
   \c|_\infty = \q \ .
\end{equation}
There must be some point $z=z_0$ in $\CC$ where this non-trivial
holonomy in $T$ untwists, at which point $T=0$. We can choose this to
be the origin $z=0$. In general this means we can globally gauge $T$
to the form  %
\begin{equation}
   T = f(r)e^{i\q}
\label{eq:chi}
\end{equation}
such that $f(0)=0$. 

The form of $f$ can be determined by subtracting the vortex
equations (\ref{eq:vortex}). We have
\begin{equation}
\begin{split}
   ig^{z\bar{z}}F^1_{z\bar{z}} - ig^{z\bar{z}}F^2_{z\bar{z}} 
      &= i\partial\left(A^1_{\bar{z}} - A^2_{\bar{z}}\right) 
           - i\bar{\partial}\left(A^1_z - A^2_z\right) \\
      &= \partial\bar{\partial} \ln f^2 \ ,
\end{split}
\label{eq:diff}
\end{equation}
so the difference of the vortex equations  (\ref{eq:vortex}) reads
\begin{equation}
   \partial\bar{\partial} \ln f^2 + 2f^2 = 2\p(\t_1-\t_2) = 2\a^2 \ .
\label{eq:feq}
\end{equation}
Finding the solution of this equation such that $f(0)=0$ and
$f(\infty)=\a$ then completely determines the vortex solution up to
gauge transformations. In particular, the individual vortex equations
give
\begin{equation}
\begin{aligned}
   F^1_{z\bar{z}} &= i\left(f^2 - \a^2\right) -2\p i \\
   F^2_{z\bar{z}} &= -i\left(f^2 - \a^2\right) 
\end{aligned}
\label{eq:F1F2}
\end{equation}
for the two field strengths. 

We see that at infinity $|T|=\a$ and the tachyon is at the minimum of
its potential. However, as one moves around the $S^1$ at infinity the
phase of the tachyon rotates giving a vortex. The vortex untwists at
the origin where $T=0$. This is the position of the D0-brane. In
general there is a modulus to move the D0-brane to at any point in the
complex plane. 

We can do the same analysis without referring directly to the vortex
equations. On $\PP^1$ we have $E_1=\mathcal{O}(1)$ and
$E_2=\mathcal{O}$ the trivial bundle. Thus we have the triple 
\begin{equation}
   \mathcal{O} \stackrel{T}{\longrightarrow} \mathcal{O}(1) \ .
\label{eq:complexeg}
\end{equation}
We can think of this as a map between holomorphic functions on $\PP^1$
(that is constant functions) to meromorphic functions with, at most, a
single pole (at the zero-brane). However, such maps lie in an exact
sequence 
\begin{equation}
   0 \longrightarrow \mathcal{O} 
       \stackrel{T}{\longrightarrow} \mathcal{O}(1)
       \longrightarrow \mathcal{O}_p \longrightarrow 0 \ .
\label{eq:equiv}
\end{equation}
That is the kernel of $T$ is zero. The cokernel, meanwhile, is simply
the sheaf of functions localized at a point $\mathcal{O}_p$. This is
precisely the set of points where $T$ vanishes. 
But since $\mathcal{O}_p$ is localized on a point $p$ it is precisely the
description of a D0-brane on $p$. Thus after condensation, $E_1$ and
$E_2$ are effectively replaced by their cokernel, representing a
D0-brane. Depending on the particular choice of the map $T$, the
D0-brane lies at different points $p$ in $\PP^1$.  

It is straightforward to show that the triple (\ref{eq:complexeg})
is $\sigma$-stable in the sense of (\ref{slope}) and (\ref{stable}), since
any sub-triple has $E_2'$ zero.

\subsection{General examples of BPS branes}

We can naturally generalize the previous example to construct, via
the process of tachyon condensation, supersymmetric $(2d-2)$-branes on
a $d$ complex dimensional Calabi--Yau manifold $X$.
Such branes are described by sheaves localized on a holomorphic
hypersurface $C$ in $X$ \cite{BBS,BSV,OOY,hm}. 
As above let us assume that $E_2$ is the trivial $U(1)$ bundle
$\mathcal{O}_X$. We then require $c_1(E_1)=[C]$, the class of
$C$. This can be achieved by 
taking $E_1$ to be the bundle $\mathcal{O}_X(C)$. Note that in general
this bundle also induces lower-dimensional brane charges. Then, for
any map $T$ we have the exact sequence
\begin{equation}
   0 \longrightarrow \mathcal{O}_X 
        \stackrel{T}{\longrightarrow} \mathcal{O}_X(C)
        \longrightarrow \mathcal{O}_C(C) \longrightarrow 0 \ ,
\end{equation}
where $\mathcal{O}_C(C)$ is a sheaf localized on $C$. As in the
previous example, this means we can replace the triple
$E_2 \stackrel{T}{\longrightarrow} E_1$ with the sheaf
$\mathcal{O}_C(C)$. Since this represents a bundle localized on $C$,
it describes a supersymmetric $(2d-2)$-brane configuration as required. 

Again, it is easy to see that the triple is $\sigma$-stable since any
sub-triple has $E_2'$ zero.

\section{BPS bound states as stable triples}

In this section we will discuss and illustrate the description of 
BPS branes as stable triples in relation with other
existing descriptions of BPS branes. 

\subsection{Stable sheaves}

Let us now relate our description of BPS D-branes as stable triples to
existing descriptions of BPS states. Locally, the bosonic D-brane
degrees of freedom are a bundle $E$ with gauge fields $A_m$ on the brane
together with scalars $\Phi$ living in the normal bundle to the brane
world-volume $C$ and the adjoint of $E$. If the brane is an embedding 
$C\subset X$, then  the normal bundle is generically non-trivial, and the
scalars are its sections. It has been argued that, the conditions that
the background on the brane preserves supersymmetry then lead to a set
of first-order holomorphic differential equations on $A_m$ and
$\Phi$~\cite{BSV,hm}. For instance, for a brane wrapping a holomorphic
two-cycle in K3, one has the Hitchin equations  
\begin{equation}
\begin{aligned}
   F_{z\bar{z}} &= [\Phi_z, \Phi_{\bar{z}}] \ , \\
   \bar{D}_{\bar{z}} \Phi_z &= D_z \bar{\Phi}_{\bar{z}} \ . 
\end{aligned}
\label{eq:hitchin}
\end{equation}
These conditions can be reinterpreted geometrically as a generalized
stability condition on the pair $(E,\Phi)$. 

A second description of a D-brane is as a sheaf $\cS$ on $X$. In
particular, it was conjectured in~\cite{hm} that BPS branes correspond
to ``coherent semistable'' sheaves on $X$. The coherent
condition means that $\cS$ fits into an exact sequence
\begin{equation}
   E_2 \longrightarrow  E_1 \longrightarrow \cS \longrightarrow 0 \ , 
\label{eq:coherent}
\end{equation}
where $E_1$ and $E_2$ are vector bundles (or more precisely, the
sheaves of sections of vector bundles). The semi-stability condition is
the generalization to sheaves of the geometrical condition of
stability for vector bundles. However, in contrast to the case for
vector bundles, there is no differential equation on the sheaf
corresponding to the condition of stability. From this point of view,
the requirement that the sheaves are stable is a conjecture. 

The sheaf description is related to the description in terms of fields
on the embedded brane $C$ as follows. One requires that on $C$ the
sheaf $\cS$ reduces to the vector bundle $E$. Furthermore, away from
$C$ the sheaf must be zero. Mathematically this means that the support
of $\cS$ is $C$, and the restriction of $\cS$ to $C$ is $E$, 
\begin{equation}
   \supp(\cS) = C \ , \qquad
   \cS|_C = E \ .
\label{eq:restrict}
\end{equation}
In fact, these conditions do not completely determine $\cS$. One can
show that the additional information required is precisely the
twisting of the scalar fields $\Phi$. Thus, locally, $\cS$ is
equivalent to the pair $(E,\Phi)$. In the case of $C$ being a curve
in K3 one can then see a close relation~\cite{del} between stable
sheaves $\cS$ and solutions of the local Hitchin
equations~\eqref{eq:hitchin}. However, in general, there is no
explicit justification of the requirement that $\cS$ be stable. 

\subsection{Stable triples}

We are proposing a description of BPS brane states as stable
triples. How does this relate to the sheaf and local-field
descriptions? 
Consider the large volume limit where we can neglect stringy
corrections. In this limit we can derive an important
result:
\begin{quote}
   \textit{In the large $\sigma$ limit, any stable triple will
   necessarily be without a kernel, that is, the map $T$ will be
   injective (one-to-one).} 
\end{quote}

To see this, suppose there is a kernel,
$\ker(T) \subset E_2$. By definition, there is then a non-trivial
sub-triple $T':\ker(T)\to 0$. Since $E_2$ is torsion-free, any
sub-sheaf of $E_2$ must be torsion-free. This implies that if
$\ker(T)$ is non-trivial, it must be supported on the whole of $X$. In
particular, we must have $\rk(\ker(T)) > 0$. Recall that for large
$\s$ the stability condition reduced to a condition on the
ranks~\eqref{theta}. However, for the sub-triple $T':\ker(T)\to 0$ we
have 
\begin{equation}
   n_2 n'_1 - n_1 n'_2 = -n_1 n'_2 < 0 \ ,
\end{equation} 
since $n'_1=0$, $n'_2 > 0$ and so the stability condition~\eqref{theta}
is violated. Thus any stable triple has $\ker(T)=0$. In general, it
will have a cokernel however. The fact that $\ker(T)=0$ implies that
there is always an exact sequence 
\begin{equation}
   0 \longrightarrow E_2 \stackrel{T}{\longrightarrow}  E_1 
     \longrightarrow \coker(T) \longrightarrow 0 \ . 
\label{eq:injective}
\end{equation}
Comparing with~\eqref{eq:coherent}, we see that, in the large~$\s$
limit, the coherent sheaf $\cS$ is simply the cokernel $\coker(T)$ of
the tachyon map. In particular, it will be supported on some
holomorphic subspace of $X$ (or $X$ itself). For example on K3, if $E_{1}$
and $E_{2}$ of the same rank, then the brane charge 
$c_1(E_{1})-c_1(E_{2})=c_1(\cS)$ must be effective, \textit{i.e.} the BPS
branes are realized as a sheaf $\coker(T)$ localized on a holomorphic
curve $C$. In particular, $c_1(E_1)-c_1(E_2)=n[C]$, where $n$ is
the rank of $\coker(T)$ on $C$. 

This appears to justify the conjecture that D-branes are described by
stable coherent sheaves. However, it turns out that the notion of
stability for an injective triple (with $T$ injective), is different
from the notion of stability of the cokernel $\cS=\coker(T)$,
considered either as a torsion sheaf on $X$ or as a vector bundle on
its support. The reason being that the $\sigma$-slope (\ref{slope}) in
the $\sigma$-stability of the triple (\ref{stable}) involves only the
ranks and the first Chern classes of $E_{1}$ and $E_{2}$. On the other
hand, the $\mu$-slope (\ref{eq:slope}) in the $\mu$-stability of the
cokernel $\cS$ involves its first Chern class. The latter is related
to the second Chern classes of $E_{1}$ and $E_{2}$, which do not enter
the $\sigma$-slope stability. 

One might expect that the vortex equations receive corrections
involving higher-order Chern classes, which could correct this
discrepancy. However, in fact, there is a basic difference between the
stability of the triple and other notions of D-branes stability.
In general, one considers the charges of a brane-antibrane system as
elements in K-theory, and searches for geometric objects that
correspond to these charges. In this paper, we suggest a representation
of the brane-antibrane charges by holomorphic triples, which satisfy
the vortex equations. For an injective triple, the K-theory charges are
then related to the Chern classes of $\coker(T)$, the difference of
$\ch(E_1)$ and $\ch(E_2)$. However, this proposal implies that the
stability condition does not depend only on the K-theory classes of a
complex and its sub-complexes, but rather on the individual
terms\footnote{We thank M. Douglas for this comment.}. The
$\sigma$-slope is expressed in terms of the sums of the ranks and
degrees of the individual terms in the complex, rather than their
alternating differences. 
%This differs, for instance, from
%$\pi$-stability \cite{d10}, where the stability of a complex depends
%on its K-theory class and those of its sub-complexes, and does not
%involve the individual terms. 

As we will discuss in the next section, we can also consider more
general webs of vector bundles. Basically we are working with the
representation of some quiver in the category of vector bundles on
$X$. We fix the `shape' of the web of bundles and
maps and then look for stability of all configurations of bundles and
maps for that shape. Thus, for instance, for vector bundles we take the
quiver of type $A_{1}$, for stable triples we take the quiver $A_{2}$
and for a sequence of $n$ vector bundles we will take the quiver
$A_{n}$. For all such objects one can define stability, but there are 
many notions of slope and stability (depending on discrete and
continuous parameters) \cite{King1,rud}\footnote{For the application 
of \cite{King1} to the stability analysis in orbifold points
see \cite{d10} section 7.}. All of these stability notions specialize
to the ordinary slope stability of vector bundles when working with
$A_{1}$. However already for $A_{2}$ there are many different
stabilities. For example there is one continuous family of stabilities
(depending on $\sigma$ or $\tau$) that we use. 

Finally, note that since there are corrections to the effective action
(\ref{eq:action}), we expect the vortex equations to be deformed. This
deformation is likely to influence the stability notion when the
corrections are not negligible, and in particular in the finite
$\sigma$ regime.

\subsection{Non-injective tachyons}

Away from the large~$\s$ limit, an interesting new possibility
arises: there may be stable triples with non-injective $T$. While we
might expect corrections in this regime, let us, for now neglect them and
discuss this possibility in more detail.  

Suppose first that $n_1=n_2=1$, \textit{i.e.} $E_{1}$ and $E_{2}$ are
global line bundles on $X$. In this case there are two possibilities:
either $T$ is the zero map or $T$ is injective. Thus in this case no
new states appear. To show this note, as above, that that $\ker(T)$
must be a sub-sheaf in $E_{2}$ and in particular is torsion free, since
$E_{2}$ has no torsion. Its rank is smaller or equal to $n_2= 1$. Now,
if $\rk(\ker(T)) = 0$, then $\ker(T) = 0$ since it cannot be supported
on any sub-variety. If $\rk(\ker(T)) = 1$, then $\ker(T) = E_{2}$ since
in this case we will have $E_{2}/\ker(T) \subset E_{1}$ is supported
on a sub-variety and hence must be zero since $E_{1}$ has no torsion. 

Consider next a more general case, let  $n_1=n_2\neq 1$. In this case
we can have solutions to the vortex equations, \textit{i.e.} stable
triples, with a non-injective tachyon. As an example, consider the
following. Let $X \subset \PP^3$ be a general quartic K3. Then the
Picard group $\Pic(X) = \mathbb{Z}$ is generated by  the hyperplane
line bundle $h=\cO_{\PP^3|X}(1)$. The tangent bundle $T_{X}$ of $X$ is
$h$-stable. In particular all line sub-bundles of $T_{X}$ will be of
the form $h^{-a}$ for some positive integer $a$. Let $a$ be the
minimal positive integer, for which $h^{-a} \subset T_{X}$. Then we
have a short exact sequence on $X$: 
\[
0 \longrightarrow h^{-a} \longrightarrow T_{X} \longrightarrow 
    h^{a}\otimes I_{Z} \longrightarrow0 \ ,
\]
where $Z \subset X$ is a finite set consisting of $24+4a^2$
 points and
$I_{Z}$ is the ideal sheaf of $Z$. Since $I_{Z} \subset {\mathcal
O}_{X}$, we get a natural inclusion $h^{a}\otimes I_{Z} \subset h^{a}$
and so we get a composition map 
\[
T: T_{X} \to h^{a}\otimes I_{Z} \subset h^{a} \subset T_{X}\otimes h^{2a} \ .
\]
Thus, in this example $E_2=T_{X}, E_1=T_{X}\otimes h^{2a}$ and
$n_1=n_2=2$. 
 
By definition $\ker(T) = h^{-a} \neq 0$. We have to check
that the triple $(T_{X}\otimes h^{2a}, T_{X},T)$ is
$\sigma$-stable for some choice of $\sigma$.
Recall the $\sigma$-stability condition (\ref{slope}), (\ref{stable}).   
We will work with the polarization $h$ on $X$. The
corresponding slopes of the members of our triple are: 
\begin{equation}
\begin{aligned}
   \mu(T_{X}) &=
      \frac{c_{1}(T_{X})\cdot h}{2} = 0 \ , \\
   \mu(T_{X}\otimes h^{2a}) &=
      \frac{(c_{1}(T_{X}\otimes h^{2a})\cdot h}{2} = 2a h^{2} = 8a \ ,
\end{aligned}
\end{equation}
since $h\cdot h=4$, and so
\begin{equation}
   \mu_{\sigma}(T) = \frac{16a + 2 \sigma}{4} \ .
\label{phi}
\end{equation}
There is a  lower bound on $\sigma$ which is universal for any stable triple
(Proposition 3.13 \cite{Brad}):
\begin{equation}
\begin{aligned}
\sigma & > \mu(E_{1})- \mu(E_{2}) \ , \\
\sigma & > 0 \ .
\end{aligned}
\end{equation}
Here it means $\sigma > 8a$.

To test for stability we need to consider a sub-triples $(E'_1,
E'_2,T')$ such that $E_2' \subset T_{X}$, $E_1' \subset T_{X}\otimes
h^{2a}$ are (saturated) subs-heaves and $T(E_2') \subset E_1'$. The
cases when $E_2'=0$ is obvious. Let $E_2' =T_{X}$, there is then only
one non-trivial case to consider:
\begin{enumerate}
\item[(1)] $T':T_{X} \to h^{a}\otimes I_{Z}$. We need to check that
   $\mu_{\sigma}(T') < \mu_{\sigma}(T)$. We have 
   \begin{equation}
      \mu_{\sigma}(T') = \frac{4a + 2 \sigma}{3} \ .
   \label{phip}
   \end{equation}
   Comparing (\ref{phi}) and (\ref{phip}) we get the stability
   condition $\sigma < 16 a$.
\end{enumerate}

Consider next $E_2'$ of rank one. The saturation condition then
implies that $E_2'$ is a line bundle and so $E_2'= h^{-k}$ with $k
\geq a$. We have
\begin{equation}
   \mu_{\sigma}(T') = 
      \frac{-4k + c_{1}(E_1')\cdot h+\sigma}{1 + \rk(E_1')} \ .
\end{equation}

There are three cases to consider:
\begin{enumerate}
\item[(2)] $E_1' = 0$. For this to be a sub-triple we need that
   $E_2'\subset \ker(T) = h^{-a}$, which is satisfied since any
   line bundle $h^{-k}$ maps non-trivially as a sub-sheaf in $h^{-a}$,
   as long as $k \geq a$. In this case, $\mu_{\sigma}(T')=-4k+\s$ and
   so we must require $\s< 16a$. 
\item[(3)] $E_1'$ has rank one and so is a line bundle. Then $E_1' =
   h^{m}$ with $m \leq a$ and so 
   \begin{equation}
      \mu_{\sigma}(T') = \frac{-4k  + 4m + \sigma}{2} \leq 
          \frac{-4a + 4a + \sigma}{2} = \frac{1}{2}\sigma \ ,
   \end{equation}
   which, by equation~\eqref{phi} is always strictly smaller than
   $\mu_{\sigma}(T)$.
\item[(4)] $E_1' = T_{X}\otimes h^{2a}$. In this case 
   \begin{equation}
      \mu_{\sigma}(T') = \frac{-4k + 16a + \sigma}{3} \leq
          4a + \frac{1}{3}\sigma \ ,
   \end{equation}
   which is again smaller than $\mu_{\sigma}(T)=8a+\sigma/2$.
\end{enumerate}
Therefore, in conclusion, the triple $(T_X\otimes h^{2a},T_X,T)$ will
be $\sigma$-stable as long as 
\begin{equation}
   8a < \sigma < 16a \ .
\end{equation} 

Let us discuss now the large $\sigma$ limit.
From the above we see that when $\sigma > 16a$
the sub-triples  $T':T_{X} \to h^{a}\otimes I_{Z}$ and
$T': h^{-k} \to 0$
have a larger $\sigma$-slope than that of the original triple.
Therefore they will destabilize the triple in the large volume limit.
Physically it means that the vortex solution corresponding to the triple will decay
in the large volume limit.

Finally, note that if we consider the case $n_1 \neq n_2$ then 
there is an upper bound on $\sigma$ for any stable triple (Proposition
3.14 \cite{Brad}): 
\begin{equation}
\sigma < \left(1+ \frac{n_1+n_2}{|n_1-n_2|}\right)
\left(\mu(E_{1})- \mu(E_{2})\right) \ .
\end{equation}
It implies that there are no stable triples with $n_1 \neq n_2$ in the
large $\sigma$ limit. At first glance, this may appear a little odd,
since one could certainly take $E_1$ and $E_2$ to be trivial, which
should lead, after tachyon condensation, simply to $n_1-n_2$ D-branes
wrapping $X$. However, the point is that one expects such a
configuration to appear as a semi-stable rather than stable
configuration, since the $n_2$ brane-antibrane pairs which annihilate
decouple from the surviving $n_1-n_2$ branes.

\section{Generalizations and dimensional reduction}

In this section we discuss some generalizations of the brane-antibrane
system. We will also show that the theory of stable triples suggests some
curious relations between brane-antibrane configurations and wrapped
branes in higher dimensions.

\subsection{Other brane-antibrane configurations} 

There is a natural generalization of the previous discussions when we
allow the $E_1$ and $E_2$ bundles to split~\cite{sharpe}. 
Suppose the $U(n_1)$ gauge
group for $E_1$ is split into $\bigoplus_iU(n_{1,i})$ and similarly
the gauge group for $E_2$ splits into $\bigoplus_jU(n_{2,j})$. We can
then write the bundles as $E_1=\bigoplus_iE_{1,i}$ for the branes and
$E_2=\bigoplus_jE_{2,j}$ for the antibranes. The form of the vortex
equation generalizes. In particular, thus far, we assumed that the
right-hand sides of equations~\eqref{eq:vortex} were simply
proportional to the identity matrix since, in general, this was the
only constant matrix in the gauge group for $E_1$ and $E_2$. However,
if $E_1$ and $E_2$ split then we have more possibilities. 

The vortex 
equations generalize to 
\begin{equation}
\begin{aligned}
   ig^{m\bar{n}}F^{1,i}_{m\bar{n}} + e\sum_j T_{i,j}T^*_{i,j} 
       &= 2\p\t_{1,i} I_{1,i} \ , \\
   ig^{m\bar{n}}F^{2,j}_{m\bar{n}} - e\sum_i T^*_{i,j}T_{i,j} 
       &= 2\p\t_{2,j} I_{2,j} \ , \\
   \sum_j T_{i,j}T^*_{k,j} &= 0 \ , \qquad \text{for $i\neq k$,} \\
   \sum_i T^*_{i,j}T_{i,k} &= 0 \ , \qquad \text{for $j\neq k$.}
\end{aligned}
\label{eq:genvortex}
\end{equation}
Here $F^{1,i}$ and $F^{2,j}$ are the field strengths for the
$U(n_{1,i})$ branes and $U(n_{2,j})$ anti-branes
respectively. Similarly the tachyon, which is a map from $E_2$ to
$E_1$, splits into a web of maps $T_{i,j}$ between $E_{2,j}$ and
$E_{1,i}$. Corresponding to these generalized vortex equations, we
would expect there to be a generalized notion of stability for the web
of bundles and maps. 

Other generalizations are also possible. In the example above we
assumed that all the maps $T_{i,j}$ were holomorphic. In general,
however, some may be anti-holomorphic. For example, suppose we have
equal numbers of $E_{1,i}$ and $E_{2,j}$ and that the tachyon maps are
zero except for $T_{i,i}$ which is holomorphic and $T_{i+1,i}$ which is
anti-holomorphic. Relabelling, this can be written as 
the $A_n$ quiver:
\begin{equation}
   E_n \stackrel{T_{n-1}}{\longrightarrow}
   E_{n-1} \stackrel{T_{n-2}}{\longrightarrow}
   E_{n-2} \stackrel{T_{n-3}}{\longrightarrow} \dots
   \stackrel{T_1}{\longrightarrow} E_1 \ .
\label{Sharpe}
\end{equation}
The corresponding vortex equations are then given by 
\begin{equation}
\begin{aligned}
   ig^{m\bar{n}}F^i_{m\bar{n}} + e T_iT^*_i - e T^*_{i-1}T_{i-1}
       &= 2\p\t_i I_i \ , \\
   T_{i-1}T_i &= 0 \ , 
\end{aligned}
\label{eq:complexvortex}
\end{equation}
where $F^i$ is the field strength for the bundle $E_i$. The second
condition on the maps implies that the sequence forms a complex. 

In general, one can consider an arbitrary quiver such that there is a
$Z_2$-grading of nodes into branes and anti-branes and a set of
holomorphic maps which always connect brane nodes and anti-brane
nodes.  

\subsection{Dimensional reduction and stable triples} 

In the following we we will discuss an interesting relation 
between the triple $(E_1,E_2,T)$ on $X$ and a single vector bundle on
$X\times\PP^1$ \cite{Brad} and its interpretation. 

Denote the two natural projections from $X\times\PP^1$ to  $X$ and
$\PP^1$ by
\begin{equation}
   p:X\times\PP^1\to X \ , \qquad
   q:X\times\PP^1\to\PP^1 \ .
\end{equation}
Consider the holomorphic tangent bundle of $\PP^1$. This is naturally
invariant under the $SU(2)$ rotations of the sphere
$\PP^1$. Geometrically it is the bundle $H^{\otimes2}$ where $H$ is
the line bundle over $\PP^1$ with Chern class one. 

We can now consider making a dimensional reduction for a set of
D-branes wrapping the $X\times\PP^1$. However, rather than assuming
the gauge-field flux is zero, we will assume that on $\PP^1$ the field
strength is given by the $SU(2)$-invariant bundle described
above. In particular, we will take the bundle on $X\times\PP^1$ to
have the form
\begin{equation}
   F = p^*E_1 \oplus \left(p^*E_2 \otimes q^*H^{\otimes2}\right) \ ,
\label{eq:Fdef}
\end{equation}
where $E_1$ and $E_2$ are arbitrary bundles on $X$. Since
$\rk{F}=n_1+n_2$, we have $(n_1+n_2)$ $(2d+2)$-branes, while
$c_1(H^{\otimes2})=2$ implies that such a bundle also describes $2n_2$
$2d$-branes wrapping $X$.   

Without a non-trivial bundle on $\PP^1$, there would be no scalars in the
dimensional reduction, since the are no covariantly constant one-forms
on $\PP^1$. However, with the particular bundle described above, there
is a pair of covariantly constant one-forms. These form a complex
scalar in the dimensional reduction. In particular, the gauge
connection will have the form
\begin{equation}
   A = \begin{pmatrix} A_1 & T\b \\ T^*\b^* & A_2 \end{pmatrix} \ ,
\end{equation}
where $A_1$ is the $E_1$-connection on $X$, $A_2$ the $E_2$-connection
on $X$ and $\b$ and $\b^*$ are the covariantly constant one-forms on
$\PP^1$. The one-forms can be normalized such that $\b\wedge\b^*$ is
the K\"ahler form on $\PP^1$. The field $T$ and its conjugate $T^*$
are in the bifundamental of $E_1$ and $E_2$.

Performing the dimensional reduction of the
Yang-Mills action for $F$ on $\PP^1\times X$ we find precisely the
action for $F_1$ and $F_2$ with tachyon $T$ as given
in~\eqref{eq:action}, with the parameter $\a^2$  (\ref{a2}) given by
the inverse volume of $\PP^1$, 
\begin{equation}
  \alpha^2 = \frac{2\pi^2}{V_{\PP^1}} \ ,
\label{eq:Vrelation}
\end{equation}
where $V_{\PP^1}$ is the volume of the $\PP^1$. 
Thus, the theory of branes and anti-branes wrapped on $X$ is obtained
as a particular $SU(2)$-invariant reduction of D-branes on
$X\times\PP^1$. Furthermore, it can be
shown~\cite{Brad} that the conditions of a stable triple $(E_1,E_2,T)$
are equivalent to requiring the bundle $F$ to be stable. While
the brane-antibrane configuration in general is not
supersymmetric, the brane on
$X\times\PP^1$ can be supersymmetric\footnote{Although
$X\times\PP^1$ is not a Calabi--Yau manifold, the single brane can
be supersymmetric if, for instance, $X\times\PP^1$ is holomorphic
subspace of some large Calabi--Yau manifold.}. 
In general, however, it may be that there is no supersymmetric bound
state with all the relevant brane charges. The dimensional reduction
relation implies that when the brane-antibrane configuration is
non-supersymmetric, there is no supersymmetric bound state with the
corresponding brane charges on $X\times\PP^1$.  

There is one further generalization. It has been shown 
in \cite{King} that general
$SU(2)$-invariant holomorphic bundles over $X\times\PP^1$ are in
one-to-one correspondence with a set of bundles and maps
(\ref{Sharpe}). There is a corresponding generalized notion of
stability. We would expect that this coincides with the generalized
vortex equations~\eqref{eq:complexvortex} discussed in the previous
section.  

Finally, note that from (\ref{eq:Vrelation}) we see again that 
since $\alpha^2$ goes like the volume of the 2-sphere (which is of the
string size) in the dimensional reduction, the region of validity of
the vortex equations is at large $\sigma \sim \alpha^2 \sim M_s^2$.

\vspace{5ex} {\bf Acknowledgement}:
We would like to thank M. Douglas for valuable discussions. 
Y.~Oz would like to thank Tel-Aviv university for hospitality during
part of this work. 
T.~Pantev is
supported in part by an NSF grant DMS-9800790 and by an Alfred
P. Sloan Research Fellowship.
D.~Waldram would like to thank The Enrico Fermi
Institute at The University of Chicago and the Physics Department of
The Rockefeller University for hospitality during the completion of
this work.

%%%%%%%%%%%%%%%%%%%%%%%%%%%%%%%%%%%%%%%%%%%%%%%%%%%%%%%%%%%%%%%%%%%%%%%%%%


\newpage 

\begin{thebibliography}{99}



\bibitem{review} A. Sen, ``Non-BPS states and Branes in String
   Theory'', APCTP winter school lectures, hep-th/9904207,
   A. Lerda and  R. Russo, ``Stable non-BPS states in string theory: a
   pedagogical review'', {\em Int.J.Mod.Phys.} {\bf A15} (2000) 771,
   hep-th/9905006; J. Schwarz, ``TASI Lectures on Non-BPS D-Brane Systems'',
hep-th/9908144.




\bibitem{1} A. Sen, ``Tachyon Condensation on the Brane Antibrane System'',
{\em  JHEP} {\bf 9808} (1998) 012.

\bibitem{Witten} E. Witten, ``D-Branes And K-Theory'', {\em JHEP} 
   {\bf 9812} (1998) 01, hep-th/9810188. 

\bibitem{horava} P. Horava, ``Type IIA D-branes, K-Theory and Matrix
   Theory'', {\em Adv. Theor. Math. Phys.} {\bf 2} (1999) 1373,
   hep-th/9812135. 

\bibitem{sentach} A. Sen, ``Descent Relations Among Bosonic
   D-branes'', {\em Int.J.Mod.Phys.} {\bf A14} (1999) 4061-4078,
   hep-th/9902105.


\bibitem{lerda} M. Frau, L. Gallot, A. Lerda and  P. Strigazzi, ``
   Stable non-BPS D-branes in Type I string theory'', {\em Nucl.Phys.}
   {\bf B564} (2000) 60, hep-th/9903123.
 

\bibitem{numerics} A. Sen, B. Zwiebach, ``Tachyon Condensation in
   String Field Theory'' {\em JHEP} {\bf 0003} (1999) 002; N.
   Berkovits, A. Sen and B. Zwiebach, `` Tachyon Condensation in
   Superstring Field Theory'', hep-th/0002211.

\bibitem{hori1}  O. Bergman, K. Hori and  P. Yi, ``Confinement on the
   Brane'', hep-th/0002223.


\bibitem{chi} J. A. Harvey, D. Kutasov and  E. J. Martinec,
``On the relevance of tachyons'',  hep-th/0003101.

\bibitem{ms} J. Majumder and  A. Sen,
``Vortex Pair Creation on Brane-Antibrane Pair via Marginal Deformation'',
{\em JHEP} {\bf 0006} (2000) 010, 
hep-th/0003124.


\bibitem{2} M. Naka, T. Takayanagi and  T. Uesugi,
``Boundary State Description of Tachyon Condensation'',
{\em  JHEP} {\bf 0006} (2000) 007, hep-th/0005114.


\bibitem{bmo}  P.  Brax, G. Mandal and  Y. Oz, ``Supergravity
Description of Non-BPS Branes'',  hep-th/0005242.



\bibitem{d1} I. Brunner, M. Douglas, A. Lawrence and
 C. R\"omelsberger, , `` D-branes on the Quintic'',
{\em JHEP} {\bf  0008} (2000) 015,
hep-th/9906200.

\bibitem{d2} D.-E. Diaconescu and J. Gomis, ``Fractional branes and 
boundary states in orbifold theories,'' hep-th/9906242



\bibitem{d3} D.-E. Diaconescu and 
C. R\"omeslberger, ``D-branes and bundles on elliptic fibrations,'' 
hep-th/9910172. 

\bibitem{d4}
 E. Scheidegger, 
``D-branes on some one and two-parameter Calabi-Yau hypersurfaces,'' 
hep-th/9912188. 

\bibitem{d5}  P. Kaste, W. Lerche, C. A. L\"utken and 
J. Walcher, ``D-branes on K3 fibrations,'' hep-th/9912147.


\bibitem{d6} 
J. Walcher, ``D-branes on K3 fibrations,'' hep-th/9912147. 

\bibitem{d7} S. Kachru, S. Katz, A. Lawrence and J. McGreevy, ``Open string
instantons and superpotentials,'' hep-th/9912151, 
{\em Phys. Rev.} {\bf D 62} (2000) 026001; ``Mirror symmetry for 
open strings,'' hep-th/0006047

\bibitem{d8}  M. Naka and M. Nozaki, ``Boundary states in Gepner models,'' 
hep-th/0001037, JHEP {\bf 0005} (2000) 027.


\bibitem{d9} I. Brunner and V. Schomerus, 
``D-branes at singular curves of Calabi-Yau 
manifolds,'' hep-th/0001132, {\em JHEP} {\bf 0004} (2000) 020;
``On Superpotentials for D-Branes in Gepner Models'', 
hep-th/0008194.



\bibitem{d10} M. Douglas, B. Fiol and  C. R\"omelsberger,,
``Stability and BPS branes'', 
hep-th/0002037.


\bibitem{d11} S. Govindarajan and T. Jayaraman, ``On the Landau-Ginzburg 
description of boundary CFTs and special Lagrangian submanifolds'' 
hep-th/0003242.



\bibitem{d12}  M. Douglas, B. Fiol and  C.  C. R\"omelsberger,,
``The spectrum of BPS branes on a noncompact Calabi-Yau'',
hep-th/0003263.


\bibitem{d13} B. Fiol and  M. Marino,
``BPS states and algebras from quivers'', 
{\em JHEP} {\bf  0007} (2000) 031,
hep-th/0006189.


\bibitem{d14} D. Diaconescu and  M. Douglas,
``D-branes on Stringy Calabi-Yau Manifolds'',
hep-th/0006224.







\bibitem{Brad}  S. Bradlow and O. Garc\'ia-Prada, ``Stable triples,
   equivariant bundles and dimensional reduction''. {\em  Math. Ann.}
   {\bf 304} (1996) 225, alg-geom/9401008;
   S. Bradlow, J. Glazebrook and  F. Kamber, ``A New look at the
   vortex equations and dimensional reduction'', alg-geom/9703019.



\bibitem{hm} J. A. Harvey and  G. Moore,
   ``On the algebras of BPS states'',  {\em Commun.Math.Phys.} {\bf
   197} (1998) 489, hep-th/9609017;

\bibitem{ghm} M. Green, J. A. Harvey and  G. Moore, ``I-Brane Inflow and
   Anomalous Couplings on D-Branes'', {\em Class.Quant.Grav.} {\bf 14}
   (1997) 47, hep-th/9605033.

\bibitem{sharpe}  E. R. Sharpe, ``D-Branes, Derived Categories, and
   Grothendieck Groups'', {\em  Nucl.Phys.} {\bf  B561} (1999) 433,
   hep-th/9902116. 


\bibitem{GM} S. Gelfand and Y. Manin, 
``Methods of Homological Algebra'', Springer (1996).

\bibitem{thomas} R. P. Thomas,
``Derived categories for the working mathematician'', math.AG/0001045.


\bibitem{Berg} A. Sen ``Supersymmetric World-Volume Action for Non-BPS
  Branes'', {\em JHEP} 9910 (1999) 008; 
  M.R. Garousi, ``Tachyon couplings on non-BPS D-branes and Dirac-Born-Infeld
  action'', hep-th/0003122; E. A. Bergshoeff, M. de Roo,
  T. C. de Wit, E. Eyras and S. Panda, ``T-Duality and Actions for
  Non-Bps D-branes'', hep-th/0003221.

\bibitem{wip} Work in progress.

\bibitem{quillen} D.~Quillen, ``Superconnnections and the Chern
   Character'', {\em Topology} {\bf 3} (1985) 89.

\bibitem{roep} G.~Roepstorff, ``Superconnections and the Higgs
   field'', {\em J.\ Math.\ Phys.}  {\bf 40} (1999) 2698,
   hep-th/9801040. 

\bibitem{BBS} K. Becker, M. Becker and  A. Strominger, ``Fivebranes,
   Membranes and Non-Perturbative String Theory'', {\em Nucl.Phys.}
   {\bf B456} (1995) 130,  hep-th/9507158. 

\bibitem{BSV}  M. Bershadsky, V. Sadov and  C. Vafa, ``D-Branes and
   Topological Field Theories'', {\em Nucl.Phys.} {\bf B463}  (1996)
   420, hep-th/9511222. 

\bibitem{OOY}  H. Ooguri, Y. Oz and  Z. Yin, `` D-Branes on Calabi-Yau
   Spaces and Their Mirrors'', {\em  Nucl.Phys.} {\bf  B477} (1996)
   407, hep-th/9606112.







\bibitem{del} R.~Donagi, L.~Ein, R.~Lazarsfeld, ``A Non-Linear
   Deformation of the Hitchin Dynamical System,'' alg-geom/9504017. 

\bibitem{King1} A. D. King, ``Moduli of
   representations of finite-dimensional algebras'',
   {\em Quart. J. Math.} Oxford {\bf 45} (1994) 515.

\bibitem{rud}   A. Rudakov, ``Stability for an abelian category'', 
{\em J. Algebra} {\bf 197}
   (1997)  231.

\bibitem{King} Reference to A. King, ``$Q$-bundles on Riemann
   surfaces,'' private communication, in~\cite{Brad}.




\end{thebibliography}
\end{document}